\documentclass[10pt]{iopart}
\usepackage[latin9]{inputenc}
\usepackage{mdframed}
\usepackage{color}
\expandafter\let\csname equation*\endcsname\relax
\expandafter\let\csname endequation*\endcsname\relax
\usepackage{amsmath}
\usepackage{amssymb}
\usepackage{graphicx}
\usepackage{esint}
\usepackage{xcolor}
\usepackage[font=bf]{caption}
\makeatletter

\usepackage{rotating}\usepackage{braket}\usepackage{amsthm}\usepackage{bbm}

\definecolor{mygreen}{rgb}{0,0.5,0}
\definecolor{myblue}{rgb}{0,0,0.75}
\definecolor{mymagenta}{cmyk}{0,1,0,0.12}

\usepackage{umoline}

\newcommand{\thickhline}{
    \noalign {\ifnum 0=`}\fi \hrule height 1pt
    \futurelet \reserved@a \@xhline
}

\makeatother

\begin{document}
\title{Quantum simulation of the dynamical Casimir effect with trapped ions}
\author{N Trautmann$^1$, P Hauke$^{2,3}$}
\address{$^1$ Institute for Applied Physics, Technical  University of
Darmstadt, 64289 Darmstadt, Germany}
\address{$^2$Institute for Theoretical Physics, University of Innsbruck, A-6020 Innsbruck, Austria}
\address{$^3$Institute for Quantum Optics and Quantum Information of the Austrian Academy of Sciences, A-6020 Innsbruck, Austria}
\ead{nils.trautmann@physik.tu-darmstadt.de}
\begin{abstract}
Quantum vacuum fluctuations are a direct manifestation of Heisenberg's uncertainty principle. 
The dynamical Casimir effect allows for the observation of these vacuum fluctuations by turning them into real, observable photons. 
However, the observation of this effect in a cavity QED experiment would require the rapid variation of the length of a cavity with relativistic velocities, a daunting challenge. 
Here, we propose a quantum simulation of the dynamical Casimir effect using an ion chain confined in a segmented ion trap.
We derive a discrete model that enables us to map the dynamics of the multimode radiation field inside a variable-length cavity to radial phonons of the ion crystal.  
We perform a numerical study comparing the ion-chain quantum simulation under realistic experimental parameters to an ideal Fabry--Perot cavity, demonstrating the viability of the mapping.  
The proposed quantum simulator, therefore, allows for probing the photon (respectively phonon) production caused by the dynamical Casimir effect on the single photon level.
\end{abstract}
\date{\today}

\maketitle

\section{Introduction}

Vacuum fluctuations lie at the heart of quantum mechanics and quantum
field theory, and many interesting physical phenomena are directly
connected to virtual photons of the vacuum, like for example the Lamb
shift \cite{Lamb_experiment} or the Casimir effect \cite{casimir1948attraction}.
The dynamical Casimir effect (DCE) \cite{moore1970quantum}, which is related to the Unruh effect and the Hawking radiation \cite{crispino2008unruh}, offers the possibility
to turn these virtual photons into real measurable photons by moving the boundaries
of a cavity with relativistic velocities and high accelerations [see \fref{fig:Cavity} (a)]. 
Such extreme velocities, however, make it difficult to observe the DCE in a cavity QED experiment. 
Several proposals have been made to overcome this problem \cite{dodonov2010current},
for example, by replacing the moving mirrors by a rapid modulation of the electrical
properties of the medium inside the cavity. 
One proposal, based on superconducting circuits \cite{johansson2010dynamical,johansson2009dynamical},
has been implemented recently \cite{wilson2011observation,dalvit2011quantum,lahteenmaki2013dynamical}.
However, in that architecture it remains a challenge to analyze the generated microwave radiation on the single photon level \cite{johansson2010dynamical}. 

In this article, we investigate the possibility to implement a quantum simulation of the DCE by using an ion chain confined
in a segmented surface trap \cite{chiaverini2005surface}, as depicted in \fref{fig:Cavity} (b) (a segmented Paul trap is
also suitable \cite{schulz2008sideband,kaufmann2012thick}). 
Hereby, the photons are mapped on the phonons of the radial vibrational modes of the ion crystal. A spatial, respectively temporal, dependence of the radial trapping potential mimics the location, respectively time modulation, of the cavity mirrors. 
The use of ion phonon modes in designed trap potentials has already been proposed for the quantum simulation of a large variety of physical
phenomena, including Bose--Hubbard-like models \cite{Porras2004b,Ivanov2009,Bermudez2010a,Bermudez2011a}
and microscopic models of friction \cite{Pruttivarasin2011,Benassi2011}. 
The dynamics of phonons moreover allows one to study the transport of heat in quantum systems \cite{LinDuan2011,Bermudez2013},
as shown experimentally in \cite{Ramm2014}.
Various laboratories~\cite{Pyka2013,Ulm2013,Mielenz2013,Ejtemaee2013} have also demonstrated that
a controlled quench of the confining potential permits the generation
of topological defects and the study of the Kibble--Zurek scenario
\cite{delCampo2010}. 

In the following, we demonstrate that this precisely controlled architecture
can also be exploited for the quantum simulation of the DCE. Using
state-of-the-art trap parameters and standard methods available for
ion traps \cite{leibfried2003quantum}, the phonons respectively photons
produced by the DCE can thus be measured on the single phonon level
with high accuracy.

In the original work on the DCE \cite{moore1970quantum}, in the following
referred to as Moore's model, the cavities were described by imposing
suitable time-dependent boundary conditions. This led to some problems
with the Hamiltonian formulation of this theory. In this article,
we will avoid these problems by introducing an appropriate model for
the propagation of the radiation field inside the mirrors. This description
can be seen as a purely phenomenological model of the mirrors that
reduces to Moore's model in a certain limit, but it can also be motivated
by microscopic considerations. The used model for the mirrors has
the additional benefit of a simple realization in the ion-trap quantum
simulator, namely by a spatial variation of the radial trapping potential.

The body of this article is divided into five parts. In
 \sref{Sec:Model_Cavity}, we introduce a Hamiltonian to
model a one-dimensional version of cavity QED with moving boundaries,
and  in \sref{Sec:Connection_models} we establish a connection between this Hamiltonian and Moore's
model. In \sref{Sec:discretizes_Hamiltonian},
we derive a discretized version of this Hamiltonian and show
how it can be mapped onto an ion chain. In \sref{Sec:numerical_comparison},
we present the results of a numerical investigation in which the ion-chain quantum simulation is compared to Moore's model using realistic experimental parameters.
Finally, in \sref{sec:experimental_considerations} we address the robustness of the simulation towards possible sources of errors and  discuss
the experimental techniques available for investigating the radiation generated by the DCE.

\begin{mdframed}[linecolor=gray]
\begin{center}
  \includegraphics[width=0.9\columnwidth]{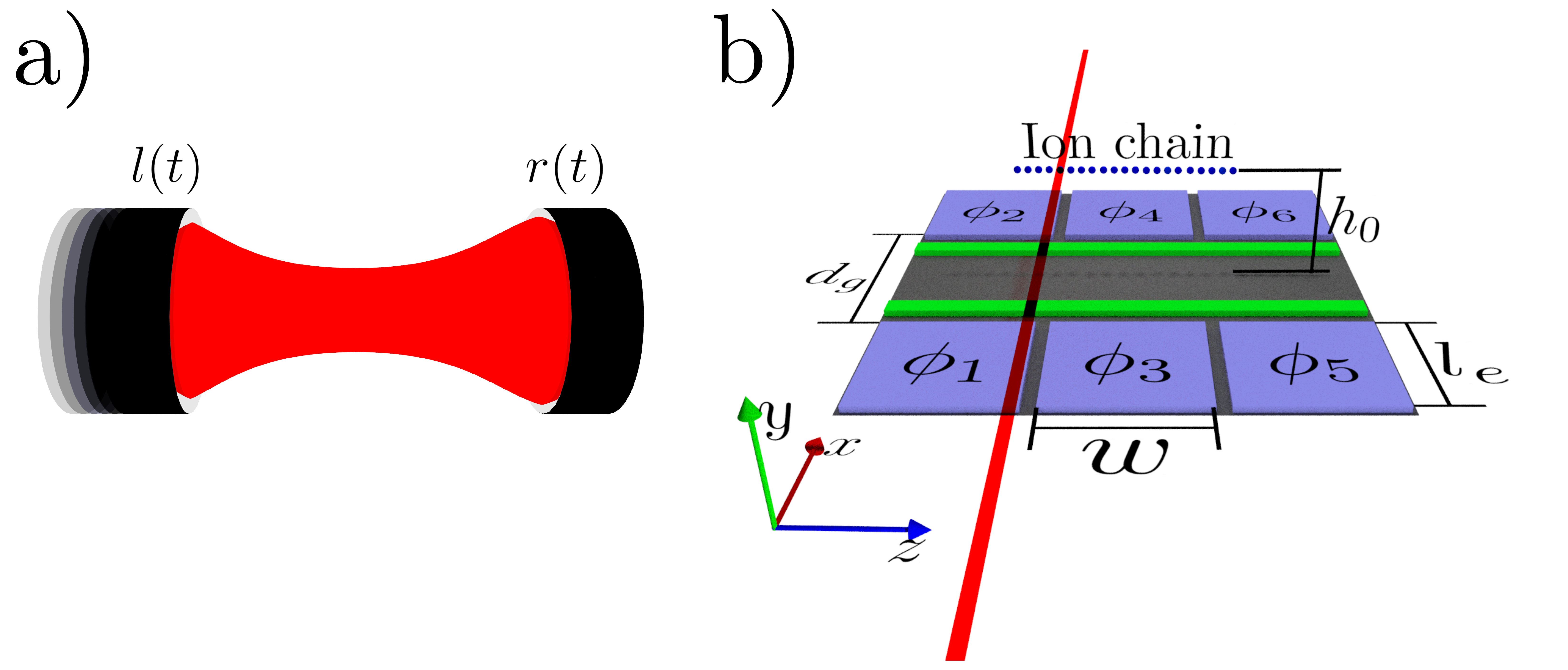}
  \captionof{figure}{\textnormal{Setup. 
(a) Dynamical Casimir effect. Modulating the positions of the left or the right mirror of a cavity [$l(t)$ and $r(t)$, respectively] results in the production of photons. 
(b) Proposed ion-trap-based quantum simulation. 
The dynamics of the radiation field inside the cavity is mapped to phonon modes of a chain of ions in a spatially-dependent trapping potential, which can be engineered in a segmented surface trap.  
The modulation of the mirror is simulated via a laser field creating a time-dependent optical trapping potential (red line in $x$-direction). 
The example shows 6 DC electrodes (blue) and two RF electrodes (green). The distance between the DC electrodes is denoted $d_{\text{g}}$, their length $l_{\text{e}}$ (both along the $x$-axis), and their width along the $z$-axis $w$. 
The ion chain (blue dots) is trapped at the center of the trap at the height $h_{0}$ above the surface. 
\label{fig:Cavity}}}
\end{center}
\end{mdframed}

\section{Model of a variable length cavity\label{Sec:Model_Cavity}}

In this section, we present a one-dimensional version of cavity QED
with moving boundaries. In order to circumvent problems connected
to the Hamiltonian formulation of the theory \cite{moore1970quantum},
we introduce a model that takes the propagation of the radiation field
inside the mirrors into account. This model can be linked to Moore's
model by considering a certain limit, but it can also be motivated
from microscopic considerations.

In the following, we consider the electromagnetic radiation field confined in a one-dimensional cavity formed by two, infinite, parallel, plane mirrors.
For the sake of simplicity, we consider only linear polarized light with an electric field oscillating along one particular axis parallel to the surface of the mirrors and we set the speed of light and all dielectric constants
equals unity, i.e., $c=\epsilon_{0}=\mu_{0}=1$.
The Hamiltonian of
our one-dimensional version of cavity QED with moving boundaries is
given by
\begin{equation}
\hat{H}=\hat{H}_{0}+\hat{H}_{1}(t)\;,\label{eq:Hamiltonian}
\end{equation}
whereby 
\begin{equation}
\hat{H}_{0}=\frac{1}{2}\int\limits _{\mathbb{R}}\hat{\Pi}^{2}(z)+\left(\frac{\partial\hat{A}}{\partial z}(z)\right)^{2}\text{d}z
\end{equation}
models the free radiation field \cite{moore1970quantum}, with $\hat{A}(z)$ being the one-dimensional
version of the vector potential and $\hat{\Pi}(z)$ being the corresponding
canonical conjugated field operator, such that the following commutation
relations hold true 
\begin{subequations}
\label{eq:canonical_relations}
\begin{eqnarray}
[\hat{A}(z_{1}),\hat{A}(z_{2})] & = & 0\;, \\
{}[\hat{\Pi}(z_{1}),\hat{\Pi}(z_{2})] & = & 0\;, \\
{}[\hat{A}(z_{1}),\hat{\Pi}(z_{2})] & = & i\hbar\delta(z_{1}-z_{2})\;.
\end{eqnarray}
\end{subequations}
The operator $\hat{H}_{1}(t)$ describes the modification of the propagation
of the electromagnetic radiation field due to the presence of the
mirror. As mentioned previously, the mirrors are not described by
imposing fixed boundary conditions but by modeling the propagation
of the field inside the mirrors. In the following, we choose 
\begin{equation}
\hat{H}_{1}(t)=\frac{1}{2}\int\limits _{\mathbb{R}}c_{1}(t,z)\hat{A}^{2}(z)\text{d}z\label{eq:H_1}
\end{equation}
with 
\begin{equation}
c_{1}(t,z)=0\;\text{for }z\in\left[l(t),r(t)\right]\label{eq:c1(t,x)}
\end{equation}
and 
\begin{equation}
c_{1}(t,z)>0\;\text{for }z\notin\left[l(t),r(t)\right]\text{\;.}
\end{equation}
As depicted in \fref{fig:Cavity} (a), the position
of the left mirror is given by $l(t)$ and the position of the right
mirror is given by $r(t)$, with $l(t)<r(t)$.

The Heisenberg equations of motion induced by the Hamiltonian $\hat{H}_{0}$
coincide with the Klein--Gordon equation of a massless particle. 
In addition to this, $\hat{H}_{1}$ introduces a time- and position-dependent effective
mass $\hbar\sqrt{c_{1}(t,z)}$. 
This effective mass induces a band-gap similar to the photonic band-gap in a photonic crystal. By choosing $c_{1}(t,z)$ sufficiently large for $z\notin\left[l(t),r(t)\right]$
the propagation of waves in that region is blocked, which models the
presence of mirrors. In the limit
$c_{1}(t,z)\rightarrow\infty$ for $z\notin\left[l(t),r(t)\right]$,
we exactly recover the dynamics of Moore's model, where the mirrors
are modeled by imposing suitable boundary conditions at $l(t)$ and $r(t)$ (see
\sref{Sec:Connection_models}). However, by taking the propagation
of the radiation field inside the mirrors into account, we circumvent
problems connected to the Hamiltonian formulation of the theory that appear
in case of Moore's model. A microscopic motivation of this Hamiltonian
can be found in \ref{Appendix_mirrors}.

It is convenient to express the field operators in the Schrödinger picture in terms of mode functions
$A_{\ell}(z)$ and $\Pi_{\ell}(z)$ and associated annihilation and creation
operators $\hat{a}_{\ell}$, $\hat{a}_{\ell}^{\dagger}$, with $\ell\in\mathbb{N}$, 
\begin{subequations}
\begin{eqnarray}
\hat{A}(z) & = & \sqrt{\frac{\hbar}{2}}\sum\limits _{\ell}(\hat{a}_{\ell}A_{\ell}(z)+\hat{a}_{\ell}^{\dagger}A_{\ell}^{*}(z))\;,\\
\hat{\Pi}(z) & = & -i\sqrt{\frac{\hbar}{2}}\sum\limits _{\ell}(\hat{a}_{\ell}\Pi_{\ell}(z)-\hat{a}_{\ell}^{\dagger}\Pi_{\ell}^{*}(z))\;.
\end{eqnarray}
\end{subequations}
In order to fulfil the canonical commutator relations in equation \eref{eq:canonical_relations}
the mode functions, which are square integrable functions for all time instances $t$, have to fulfil the conditions
\begin{subequations}
\begin{eqnarray}
\sum_{\ell}A_{\ell}(z_{1})A_{\ell}^{*}(z_{2})-\text{c.c.}&=&0\;,\\
\sum_{\ell}\Pi_{\ell}(z_{1})\Pi_{\ell}^{*}(z_{2})-\text{c.c.}&=&0\;,\\
\sum_{\ell}A_{\ell}(z_{1})\Pi_{\ell}^{*}(z_{2})+\text{c.c.}&=&2\delta(z_{1}-z_{2})\;,
\end{eqnarray}
\end{subequations}
wherein c.c.\ stands for the complex conjugate.
There is no unique choice for the mode functions, and different choices
will lead to different non-equivalent definitions of photon numbers.
In order to fix this problem, which has already been discussed in
\cite{moore1970quantum}, we will exploit that there is in fact a
canonical choice for the mode functions whenever the function
$c_{1}(t,z)$ modeling the boundaries of the cavity is time independent.
In this case, we can choose the mode functions to be solutions of
a generalized version of the Helmholtz equation 
\begin{subequations}
\label{eq:Helmholtz_all}
\begin{eqnarray}
0 & = & \left(-\omega_{\ell}^{2}-\frac{\partial^{2}}{\partial z^{2}}+c_{1}(t,z)\right)g_{\ell}(z)\label{eq:Helmholtz}\;,\\
A_{\ell}(z) & = & \frac{1}{\sqrt{\omega_{\ell}}}g_{\ell}(z)\label{eq:Initial_A_i}\;,\\
\Pi_{\ell}(z) & = & \sqrt{\omega_{\ell}}g_{\ell}(z)\label{eq:Initial_Pi_i}\;,
\end{eqnarray}
\end{subequations}
with $\omega_{\ell}>0$. The solutions $g_{\ell}(z)$ are properly normalized and orthogonal functions, 
\begin{equation}
\int_{\mathbb{R}}g_{\ell_{1}}(z)g_{\ell_{2}}^{*}(z)\text{d}z=\delta_{\ell_{1},\ell_{2}}\,,
\end{equation}
which form a complete basis of the space of square integrable functions
$L^{2}(\mathbb{R})$, i.e., 
\begin{equation}
\sum_{\ell}g_{\ell}(z_{1})g_{\ell}^{*}(z_{2})=\delta(z_{1}-z_{2})\;.
\end{equation}
For the particular choice of mode functions according to equations \eref{eq:Helmholtz}-\eref{eq:Initial_Pi_i}, the time evolution
of the field operators in the Heisenberg picture in case of fixed boundaries is just given by 
\begin{subequations}
\begin{eqnarray}
\hat{A}(t,z) & = & \sqrt{\frac{\hbar}{2}}\sum\limits _{\ell}\left\{\exp\left[-i\omega_{\ell}\left(t-t_{0}\right)\right]\hat{a}_{\ell}A_{\ell}(z)+\text{H.c.}\right\}\;,\\
\hat{\Pi}(t,z) & = & - i \sqrt{\frac{\hbar}{2}}\sum\limits _{\ell}\left\{\exp\left[-i\omega_{\ell}\left(t-t_{0}\right)\right]\hat{a}_{\ell}\Pi_{\ell}(z)-\text{H.c.}\right\}\;.
\end{eqnarray}
\end{subequations}
By this choice of mode functions, we obtain a canonical definition for the photon numbers. 

In order to discuss the production of photons, we consider like Ref.~\cite{moore1970quantum}
an experiment that can be divided in three stages. In stage I, which
corresponds to the time interval $[t_{0},t_{1})$, we consider a cavity
with fixed boundaries, i.e., we assume that  throughout this time interval the function $c_{1}(t,z)$
is constant in time. In stage II, which
corresponds to the time interval $[t_{1},t_{2})$, we consider a cavity
with time-dependent boundaries, i.e., a time-dependent function $c_{1}(t,z)$.
In stage III, corresponding to the time interval $[t_{2},\infty)$,
we again consider a cavity with fixed boundaries, i.e., $c_{1}(t,z)$ is now again time independent. Hereby, $c_{1}(t,z)$
for $t>t_{2}$ and for $t<t_{1}$ do not necessarily have to coincide.

Stage I is needed in order to be able to properly define our initial
field configuration, which we choose to be the vacuum state of the
radiation field. In stage II of the experiment, the actual photon
production will take place. Finally, in stage III of the experiment,
during which the mirrors are again at rest, a suitable measurement
of the photon numbers and their distribution among the (now again well defined)
modes is performed.

\section{Connection to Moore's model\label{Sec:Connection_models}}

In his original work, Moore concluded that the quantum
theory of the radiation field in a variable length cavity possesses
no Hamiltonian and no Schr\"odinger picture (later an effective Hamiltonian which describes the essential features
of the physical processes has been derived \cite{law1994effective}). 
To quantize the radiation field, he instead exploited a certain symplectic
structure on a function space $S$, the space of classical solutions
of the wave equation. In this section, we will establish a connection
between our model and Moore's original formulation of the theory. 

For doing so, we consider the Heisenberg equations of motions for our
model induced by $\hat{H}$, 
\begin{subequations}
\label{eq:Heisenberg}
\begin{eqnarray}
\frac{\partial}{\partial t}\hat{A}(t,z) & = & \hat{\Pi}(t,z)\,,\label{eq:Heisenberg_A}\\
\frac{\partial}{\partial t}\hat{\Pi}(t,z) & = & \frac{\partial^{2}}{\partial z^{2}}\hat{A}(t,z)-c_{1}(t,z)\hat{A}(t,z)\;.\label{eq:Heisenberg_Pi}
\end{eqnarray}
\end{subequations}
It is possible to solve these equations of motion by expanding the field operators $\hat{A}(t,z)$, $\hat{\Pi}(t,z)$ using appropriate time-dependent mode functions $A_{\ell}(t,z)$, $\Pi_{\ell}(t,z)$, 
\begin{subequations}
\begin{eqnarray}
\hat{A}(t,z) & = & \sqrt{\frac{\hbar}{2}}\sum\limits _{\ell}(\hat{a}_{\ell}A_{\ell}(t,z)+\hat{a}_{\ell}^{\dagger}A_{\ell}^{*}(t,z))\;,\\
\hat{\Pi}(t,z) & = & -i\sqrt{\frac{\hbar}{2}}\sum\limits _{\ell}(\hat{a}_{\ell}\Pi_{\ell}(t,z)-\hat{a}_{\ell}^{\dagger}\Pi_{\ell}^{*}(t,z))\;.
\end{eqnarray}
\end{subequations}
The time-dependence of these mode functions is governed by the following
equations 
\begin{subequations}
\label{eq:time_mode_A_i_and_Pi_i}
\begin{eqnarray}
\left(\frac{\partial^{2}}{\partial t^{2}}-\frac{\partial^{2}}{\partial z^{2}}+c_{1}(t,z)\right)A_{\ell}(t,z) & = & 0\label{eq:time_mode_A_i}\;,\\
\frac{\partial}{\partial t}A_{\ell}(t,z)\mid_{t=t_{0}} & = & -i\Pi_{\ell}(t_{0},z)\;.\label{eq:eq:time_mode_Pi_i}
\end{eqnarray}
\end{subequations}
By using these equations to describe the time evolution of the mode functions during stage
II, we are able to establish a connection to the annihilation and
creation operators $\hat{a}_{\ell}$ and $\hat{a}_{\ell}^{\dagger}$
associated to the mode functions in stage I and stage III. Since these
are well defined by equations~ \eref{eq:Helmholtz}-\eref{eq:Initial_Pi_i},
this connection allows us to describe the photon production caused by
the moving mirrors.

We are now in the position to establish the connection to Moore's model as follows. The real and imaginary parts of the mode functions $A_{\ell}(t,z)$ correspond to functions in the vector space $S$ defined by Moore \cite{moore1970quantum},  equipped with the time-invariant symplectic form 
\begin{equation}
 \{f_{1}\mid f_{2}\} =\int\limits _{\mathbb{R}} f_{2}(z,t)\frac{\partial f_{1}(z,t)}{\partial t}-\frac{\partial f_{2}(z,t)}{\partial t}f_{1}(z,t)\mathrm{d}z\;, 
\end{equation}
which is used in \cite{moore1970quantum} to quantize the theory.

In order to connect our model with the boundary conditions
used in \cite{moore1970quantum}, we consider the limit $c_{1}(t,z)\rightarrow\infty$
for $z$ outside the interval $\left[l(t),r(t)\right]$, which corresponds
to the limit of perfectly reflecting mirrors. In this limit, those canonical
modes of stages I and III that have $\omega_{\ell}^{2}\ll c_{1}(t,z)$, $z\notin[l(t),r(t)]$, will have their main support in the region $[l(t),r(t)]$ corresponding to the actual cavity. Outside of this region, the corresponding mode functions will experience an exponential damping.
In the limit $c_{1}(t,z)\rightarrow\infty$ for $z\notin[l(t),r(t)]$, this exponential decrease becomes equivalent to the boundary conditions 
\begin{equation}
0=A_{\ell}(t,l(t))=A_{\ell}(t,r(t))\label{eq:Boundary_condition}
\end{equation}
chosen by \cite{moore1970quantum}. Similar considerations also hold
true during stage II. Thus, the dynamics of the mode functions can also be modelled by the boundary conditions \eref{eq:Boundary_condition} if $c_{1}(t,z)$ is sufficiently large for $z\notin[l(t),r(t)]$.
As a consequence, our model will lead to the same results as Moore's
model in all three stages in the limit $c_{1}(t,z)\rightarrow\infty$ for $z\notin[l(t),r(t)]$.

\section{Mapping to ion chain\label{Sec:discretizes_Hamiltonian}}

In this section, we map the dynamics induced by the Hamiltonian $\hat{H}$ {[}see equation \eref{eq:Hamiltonian}{]} onto a system of trapped ions. 
To perform the mapping, we first introduce a model that allows us to represent the continuous one-dimensional space by the discrete ion positions. For the simulation of the DCE, a central region of the ion chain will then assume the role of the space within the cavity while portions towards the ends will stand in for the mirrors.
Afterwards, we describe how the photons can be mapped onto collective radial phonon modes of the ion crystal. 

\subsection{Discretized version of the radiation-field Hamiltonian}

To perform the mapping of the Hamiltonian $\hat{H}$ to an ion chain that has discrete positions, we first need to express it in discretized variables. 
This can be achieved by dividing the real axis $\mathbb{R}$ in suitable intervals $\left[z_{j},z_{j+1}\right]$, with $z_{j}<z_{j+1}$, $j\in\mathbb{Z}$. 
For simplicity, we describe here the case of equidistant ion spacings, where the intervals are of equal length $d=z_{j}-z_{j+1}$.
It is straightforward to generalize the subsequent discussion to intervals of non-equal length. 
This permits one to take a non-equidistant distribution of the ions and a resulting variation of nearest-neighbor coupling strengths into account. 

To arrive at a discretized Hamiltonian, we introduce the coarse-grained operators 
\begin{subequations}
\begin{eqnarray}
\hat{A}_{j} & =\frac{1}{\sqrt{d}} & \int_{z_{j}}^{z_{j+1}}\hat{A}(z)\mathrm{d}z\,,\label{eq:discetized_A}\\
\hat{\Pi}_{j} & =\frac{1}{\sqrt{d}} & \int_{z_{j}}^{z_{j+1}}\hat{\Pi}(z)\mathrm{d}z\,,\label{eq:eq:discetized_Pi}
\end{eqnarray}
\end{subequations}
which fulfill the commutation relations 
\begin{subequations}
\begin{eqnarray}
[\hat{A}_{i},\hat{A}_{j}] & = & 0\;, \\
{}[\hat{\Pi}_{i},\hat{\Pi}_{j}] & = & 0\;, \\
{}[\hat{A}_{i},\hat{\Pi}_{j}] & = & i\hbar\delta_{i,j}\;.
\end{eqnarray}
\end{subequations}
By using these operators, we obtain a discretized version of the Hamiltonian $\hat{H}$ , 
\begin{equation}
\hat{H}^{\text{d}}=\hat{H}_{0}^{\text{d}}+\hat{H}_{1}^{\text{d}}(t)\,,\label{eq:Hamiltonian_discrete}
\end{equation}
with 
\begin{equation}
\hat{H}_{0}^{\text{d}}  =  \frac{1}{2}\sum_{i\in\mathbb{Z}}\left[\hat{\Pi}_{i}^{2}+d^{-2}\left(\hat{A}_{i+1}-\hat{A}_{i}\right)^{2}\right]\,,
\end{equation}
and
\begin{equation}
\hat{H}_{1}^{\text{d}}(t)  =  \frac{1}{2}\sum_{i\in\mathbb{Z}}c_{1}^{i}(t)\hat{A}_{i}^{2}\,, 
\end{equation}
where   
\begin{equation}
c_{1}^{i}(t)=\frac{1}{d}\int_{z_{i}}^{z_{i+1}}c_{1}(t,z)\mathrm{d}z\;.\label{eq:c_1^j}
\end{equation}
In the above, we assumed that the modes of most importance are slowly varying
on the length scale induced by the interval lengths $d$,
which is well satisfied for the low-energetic modes. In the limit
$d\rightarrow0$, we recover the dynamics induced by the original
Hamiltonian $\hat{H}$.

\subsection{Implementation using radial phonons of an ion chain}

To implement the Hamiltonian $\hat{H}^{\text{d}}$, we map it to the
radial motion of a linear ion chain. Hereby, the position and momentum
of each ion represent, respectively, the fields $\hat{A}(z)$ and $\hat{\Pi}(z)$ 
averaged over one of the intervals $\left[z_{i},z_{i+1}\right]$.
We consider a linear chain of $N$ ions, confined
in a suitable trapping potential $V_{\text{trap}}$ and with equilibrium
positions $\mathbf{R}_{1},\mathbf{R}_{2},\dots,\mathbf{R}_{N}$. If
we assume that the deviations from the equilibrium positions are small,
we can apply a second-order tailor expansion around the equilibrium
positions. In this harmonic approximation, the motional degrees of
freedom along different symmetry directions are uncoupled. In the
following, we focus on the radial motion along the $x$-direction,
which is described by the Hamiltonian 
\begin{eqnarray}
\hat{H}^{\text{ions}} & = & \frac{1}{2m}\sum\limits _{i=1}^{N}\hat{P}_{i}^{2}-\frac{Z^{2}e^{2}}{8\pi\epsilon_{0}}\sum\limits _{i>j}\frac{(\hat{X}_{i}-\hat{X}_{j})^{2}}{\mid\mid\mathbf{R}_{i}-\mathbf{R}_{j}\mid\mid^{3}}\nonumber \\
 & + & \frac{1}{2}\sum\limits _{i=1}^{N}\frac{\partial^{2}V_{\text{trap}}(t,\mathbf{R}_{i})}{\partial x^{2}}\hat{X}_{i}^{2}\;.
\end{eqnarray}
By applying the canonical transformation 
\begin{subequations}
\begin{eqnarray}
\hat{X}_{i} & \rightarrow & (-1)^{i}\hat{X}_{i}\;,\\
\hat{P}_{i} & \rightarrow & (-1)^{i}\hat{P}_{i}\;,
\end{eqnarray}
\end{subequations}
we obtain 
\begin{eqnarray}
\hat{H}^{\text{ions}} & = & \frac{1}{2m}\sum\limits _{i=1}^{N}\hat{P}_{i}^{2}-\frac{1}{2}\sum_{i>j}{(-1)^{i-j}}k_{i,j}(\hat{X}_{i}-\hat{X}_{j})^{2}\nonumber \\
 & + & \frac{1}{2}\sum\limits _{i=1}^{N}\chi_{i}(t)\hat{X}_{i}^{2}\;,\label{eq:Hions}
\end{eqnarray}
with 
\begin{eqnarray}
k_{i,j} & = & \frac{Z^{2}e^{2}}{4\pi\epsilon_{0}}\mid\mid\mathbf{R}_{i}-\mathbf{R}_{j}\mid\mid^{-3}\;,\\
\chi_{i}(t) & = & \frac{\partial^{2}V_{\text{trap}}(t,R_{i})}{\partial x^{2}}-\sum_{j\neq i}(1-(-1)^{i-j})k_{i,j}\;.\label{eq:chi_i}
\end{eqnarray}

The mapping of the dynamics of the radiation field in a variable-length
cavity to the dynamics of the ion chain is achieved by the formal similarity
of equation~\eref{eq:Hions} and equation~\eref{eq:Hamiltonian_discrete}.
If we restrict the phonon Hamiltonian to nearest-neighbor interactions and consider an equidistant
distribution of the ions
$\hat{H}^{\text{ions}}$ indeed reproduces the Hamilton $\hat{H}^{\text{d}}$,
if we establish the following relations 
\begin{subequations}
\label{eqs:rel_all}
\begin{eqnarray}
 & \hat{\Pi}_{i}=\hat{P}_{i}/\sqrt{m}\;,\label{eq:rel_Pi_P}\\
 & \hat{A}_{i}=\hat{X}_{i}\sqrt{m}\;,\label{eq:rel_A_X}\\
 & d^{-2}=k_{i+1,i}/m\;,\label{eq:Rel_J_d}\\
 & c_{1}^{i}(t)=\chi_{i}(t)/m\;.\label{eq:chi_c}
\end{eqnarray}
\end{subequations}
The translation table from simulated objects to simulating objects is summarized in \tref{tab:Translation-tabel}. 
\begin{table}
\caption{
\textnormal{Summary of the connection between the simulated objects (radiation
field in variable-length cavity) and the simulating objects (ion chain
with time-dependent trapping potential) \label{tab:Translation-tabel}}}
\begin{tabular}{l l}

\thickhline
\vspace*{0.2cm}
{Simulated Objects}  & {Simulating Objects }
\tabularnewline
\hline

{Photons}  & {Phonons}\vspace*{0.3cm}\tabularnewline
{Field operators}  & {Position and Momentum operators }\tabularnewline
\vspace*{0.3cm}
{$\hat{A}_{i}$ and $\hat{\Pi}_{i}$}  & {of the radial motion $\hat{X}_{i}$ and $\hat{P}_{i}$ }\tabularnewline
{Variable-length cavity\hspace{0.1cm}}  & {Spatial- and time-dependent trap-}\tabularnewline
{modeled by $c_{1}^{i}(t)$}  & {ping potential $V_{\text{trap}}$, respectively $\chi_{i}$}\vspace*{0.2cm}\tabularnewline
{discretized \hspace{0.1cm}}  & {Hamiltonian $\hat{H}^{\text{ions}}$ describing the}\tabularnewline
{Hamiltonian $\hat{H}^{\text{d}}$}  & {radial motion of the ion chain}\vspace*{0.2cm}\tabularnewline
\thickhline
\end{tabular}
\end{table}

The additional interaction terms beyond nearest neighbors could, to
a large extent, be reabsorbed in a different choice for the approximations
leading to $\hat{H}^{\text{d}}$. In fact, the discretization $\frac{1}{2}\sum_{j\in\mathbb{Z}}d^{-2}\left(\hat{A}_{j+1}-\hat{A}_{j}\right)^{2}$
of the term $\frac{1}{2}\int\limits _{\mathbb{R}}\left(\frac{\partial\hat{A}}{\partial z}(z)\right)^{2}\mathrm{d}z$
is not unique. The chosen discretization is motivated by the following
approximation of the first derivative 
\begin{equation}
\frac{\partial\hat{A}}{\partial z}(z)\approx\frac{1}{d}\left[\hat{A}(z+d)-\hat{A}(z)\right]\,.
\end{equation}
There are, however, other possible approximations, and as such other
possible versions of $\hat{H}^{\mathrm{d}}$ (which give proper results if
$d$ is sufficiently small). In principle, this liberty could be exploited
to account for interaction terms beyond nearest neighbors, but we
will see below in \sref{Sec:numerical_comparison} that this
is not necessary to get good qualitative and even quantitative agreement
using realistic parameters. The underlying reason is the fast decrease
of dipolar interactions with distance. Since the integral over  dipolar interactions
in one dimension converges, any perturbative effect due to interactions beyond nearest neighbors will saturate
quickly when increasing the system size.

The remaining ingredient to simulate the DCE is the proper choice
of the coefficients $\chi_{i}$. By virtue of equations~\eref{eq:chi_c}
and~\eref{eq:c_1^j}, these are directly connected to the function
$c_{1}(t,z)$. Since the outermost ions represent the mirrors while
the central part of the ion chain is to simulate the vacuum within
the cavity, the coefficients $\chi_{i}$ need to vary across the chain.
This can be achieved by exploiting that the $\chi_{i}$ are not only
determined by the confining potential $V_{\text{trap}}$ along the
$x$-direction, but also by the Coulomb repulsion between the neighboring
ions, see equation~\eref{eq:chi_i}. As we will show in the next section,
by properly balancing both contributions, we can design suitable coefficients
$\chi_{i}$, even with a small number of electrode segments.

To induce the dynamical Casimir effect, the coefficients $\chi_{i}$, moreover,  
need to vary in time during stage II, which can be realized by a time-dependent
trapping potential $V_{\text{trap}}$. Since a small spatial motion
of the mirrors corresponds to a change of the $\chi_{i}$ only over
a small number of ions, this requires some local addressability. A
suitable modulation can be achieved by combining a time-independent
electric potential $V_{\text{E}}$ (including the RF potential), generated
via the segmented Paul trap, with an additional time-dependent optical
potential $V_{\text{O}}$ derived from a laser that addresses only
one or a few of the ions \cite{schneider2010optical}. Using the time-dependent
optical potential, we can vary the boundary of the cavity during stage
II of the experiment. This completes all required ingredients for
the simulation of the DCE. In the next section, we demonstrate that
good agreement to the ideal model can be obtained already for about
20 ions and in present-day architectures.

\section{Numerical comparison between Moore's model and ion-chain quantum
simulation \label{Sec:numerical_comparison}}

In this section, we compare the ion-chain quantum simulator for realistic experimental parameters to the idealized model introduced by Moore
\cite{moore1970quantum}. 
We first compute the trapping potential for realistic experimental parameters for a segmented trap, where for concreteness we consider a surface trap as depicted in \fref{fig:Cavity} (b), although a segmented Paul trap is equally well suited. It turns out that our requirements on the surface--ion distance or the width of the DC-electrodes are not very high and are met by many existing experimental setups, for example those of Refs.~\cite{schulz2008sideband,vittorini2013modular,ospelkaus2011microwave,allcock2012heating,brown2011coupled,stick2006ion}.
Afterwards, we present numerical results for the photon production, as can be simulated in a chain of $20$ ions with current technology.

\subsection{Trapping potential for realistic parameters}

The trap we consider consists of only $6$ DC electrodes. This turns
out to be sufficient to form a suitable electric potential $V_{\text{E}}$,
which we compute by using the framework presented in \cite{wesenberg2008electrostatics}
and by applying the gapless plane approximation. For the sake of simplicity,
we assume that the extension of the RF-electrodes along the $z$-axis
as well as the length of the DC-electrodes along the $x$-axis $l_{\text{e}}$ are infinite.
Inspired by \cite{vittorini2013modular}, we assume a possible setup
with $w=80\mu\text{m}$, $h_{0}=80\mu\text{m}$, $d_{\text{g}}=230\mu\text{m}$, and we consider singly-charged ions.
We set the voltages of the DC electrodes to the values 
\begin{eqnarray}
 & \phi_{1}=\phi_{2}=\phi_{5}=\phi_{6}=-5.61V\;,\nonumber \\
 & \phi_{3}=\phi_{4}=1.75V\;,
\end{eqnarray}
and use the RF electrodes to induce a confining potential that corresponds
to a trapping frequency 
\begin{equation}
\omega_{\text{RF}}=\sqrt{7.40\overline{k}/m}\;.
\end{equation}
Here, $\overline{k}=\frac{e^{2}}{4\pi\epsilon_{0}\overline{\Delta R}^{3}}$
is the average nearest-neighbor coupling strength, with 
\begin{equation}
\overline{\Delta R}=\mid\mid\mathbf{R}_{20}-\mathbf{R}_{1}\mid\mid/19=4.00\mu\text{m}
\end{equation}
being the average nearest-neighbor distance. For calcium ions ($^{40}\text{Ca}^{+}$), for example, we obtain 
\begin{equation}
\sqrt{\overline{k}/m}=2\pi\cdot1.17\text{ MHz}
\end{equation}
and $\omega_{\text{RF}}=2\pi\cdot3.18\text{ MHz}$.
All these values lie in the range of existing experimental setups.

The calculation of the coefficients $\chi_{i}$ for these parameters
yields the result depicted in \fref{fig_chi}. Here, we took
the non-equidistant distribution of the ions for this trapping potential
as well as all possible interactions (beyond nearest neighbors) into
account. 

\begin{mdframed}[linecolor=gray]
\begin{center}
  \includegraphics[scale=0.35]{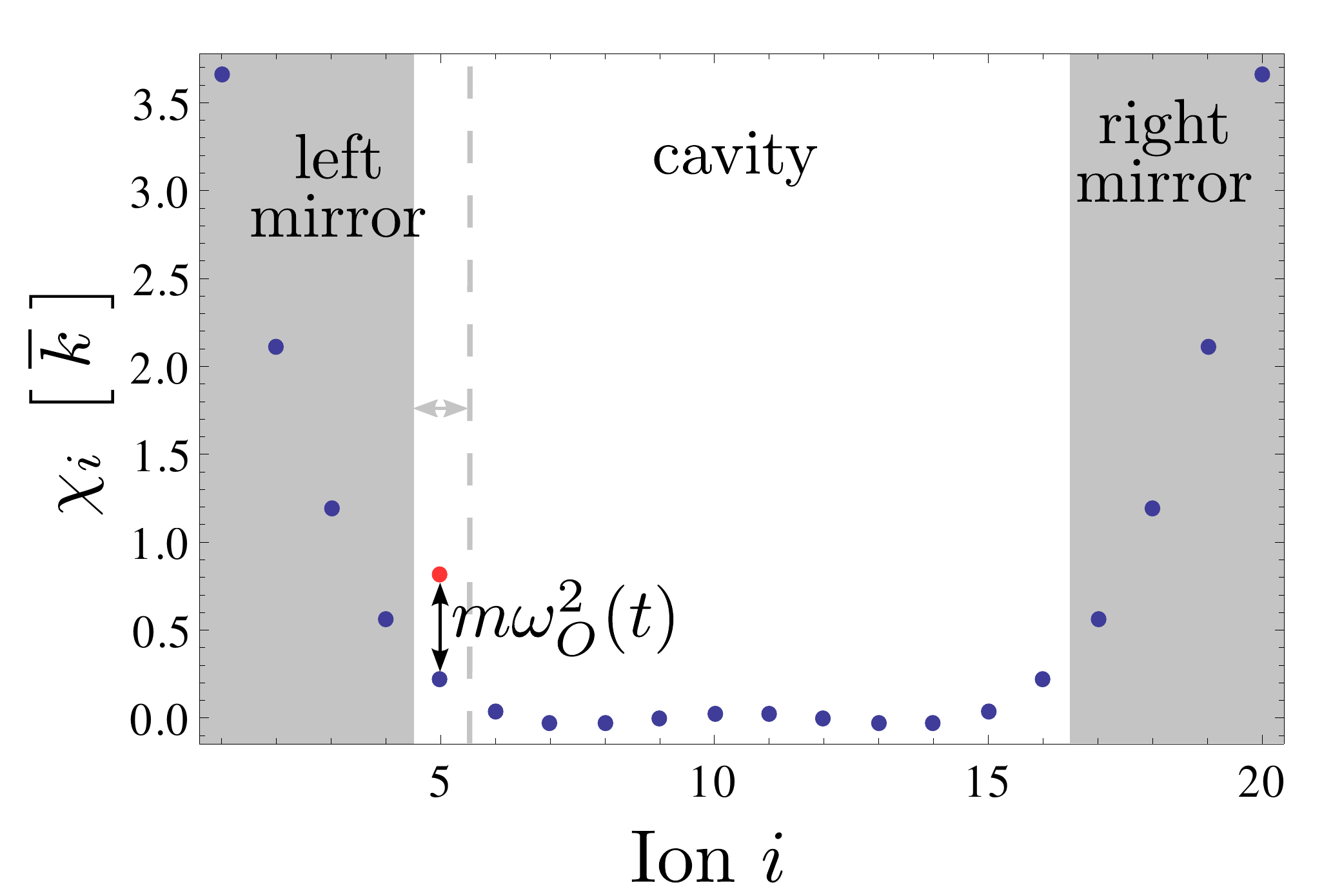}
  \captionof{figure}{\textnormal{The coefficients $\chi_{i}$ induced by the time-independent potential
$V_{\text{E}}$, plotted over the ion number.
The ions in the grey (white) areas represent the radiation
field inside the mirrors (cavity). The time dependence of the cavity
length is simulated by the time-dependent optical trapping potential
characterized by $\omega_{O}^{2}(t)$. \label{fig_chi} }}
\end{center}
\end{mdframed}

The trapping potential is adjusted such that the $\chi_{i}$ deviate
significantly from zero only in the outer regions of the chain. Since
the electric field experiences an exponential damping within the mirrors,
a rather small number of ions proves sufficient to model the space
inside the mirrors, in this example ions 1 to 4 and 17 to 20. The
field inside the cavity is represented by the inner part of the ion
chain, i.e., the ions 5 to 16.

We subject this ion chain to the three-stage protocol defined in \sref{Sec:Model_Cavity}.
The time dependence in stage II that we consider corresponds to a periodically oscillating left mirror, i.e., 
\begin{eqnarray}
 & r(t)=r_{0}\;,\label{eq:right_b-1}\\
 & l(t)=l_{0}+\delta & \begin{cases}
\begin{array}{c}
0\\
\sin^{2}(\omega_{D}(t-t_{1})/2)\\
\sin^{2}(\omega_{D}(t_{2}-t_{1})/2)
\end{array} & \begin{array}{c}
t\in[t_{0},t_{1})\\
t\in[t_{1},t_{2})\\
t\in[t_{2},\infty)
\end{array}\end{cases}\;,\nonumber \\
\end{eqnarray}
where $r_{0}$ and $l_{0}$ denote the initial positions of the right
and left mirrors, respectively. Further, $\delta/2$ is the amplitude of the variation and $\omega_{D}$ the driving frequency. This choice of the time dependence
of the boundaries leads to an efficient photon production \cite{meplan1996exponential,dodonov1998resonance}.

In order to simulate these mirror trajectories, we use a laser beam
to change the radial confinement of ion 5 such that 
\begin{equation}
\omega_{O}^{2}(t)=\alpha\frac{\overline{k}}{m}\begin{cases}
\begin{array}{c}
0\\
\sin^{2}(\omega_{D}(t-t_{1})/2)\\
\sin^{2}(\omega_{D}(t_{2}-t_{1})/2)
\end{array} & \begin{array}{c}
t\in[t_{0},t_{1})\\
t\in[t_{1},t_{2})\\
t\in[t_{2},\infty)
\end{array}\end{cases}\label{eq:optical_potential}
\end{equation}
where we choose $\alpha=0.6$. It is hereby not a strict requirement to address
precisely a single ion. Addressing several neighboring ions due to
a larger beam waist just corresponds to a larger variation $\delta$
of the position of the left mirror. Similarly, a different depth of
the optical potential $\alpha\overline{k}/m$ also just amounts to
a different $\delta$.

In the following section, we will numerically compare the dynamics for the ideally
conducting mirrors modeled according to Moore \cite{moore1970quantum}
with our ion-chain quantum simulation. For evaluating the photon number,
we choose the canonical set of mode functions for the experimental
stage III, defined by equations \eref{eq:Helmholtz}-\eref{eq:Initial_Pi_i}.
We order the mode frequencies $\omega_\ell$ as $\omega_{1}<\omega_{2}<...$ and denote with $\hat{n}_{1},\hat{n}_{2}...$
the corresponding photon number operators. For better comparison,
we choose $r_{0},l_{0}$, and $\delta$ such that the frequency of
the lowest instantaneous eigenmode in the cavity matches the frequency
of the lowest instantaneous vibrational mode of the ion chain at time
instances corresponding to the maximal and minimal cavity length,
i.e., $\omega_{D}(t-t_{1})=0$ and $\omega_{D}(t-t_{1})=\pi$. This
matching is obtained by 
\begin{eqnarray}
 r_{0}-l_{0}&=&15.22d\;,\label{eq:match_r0_l0}\\
 \delta&=&0.72d\;,\label{eq:match_delta}
\end{eqnarray}
where $d=\left({\overline{k}/m}\right)^{-1/2}$ denotes the length of the discretization
intervals $\left[z_{i},z_{i+1}\right]$ used in the previous section.
The result for the `cavity length', $r_{0}-l_{0}$, can be understood by recalling that each ion represents the averaged field in one of the intervals $\left[z_{i},z_{i+1}\right]$
and that the field inside the cavity is roughly represented by the 12 inner ions. 
The deviation between $15.22d$ and the length $12d$ expected from this simple consideration is caused by the non equidistant distribution of the ions and the discretization
of the field.

\subsection{Numerical simulation of the dynamical Casimir effect}

We have now all the necessary parameters to numerically simulate the DCE as can be studied in a realistic ion-trap experiment.  
We assume that during the experimental stage I the ion chain resides in its vibrational ground state. 
Since this initial state corresponds to the vacuum of the radiation field, all photons measured in stage III are those
that have been produced in stage II. In \fref{fig:n_without_correction-1}, we show the final photon number in the modes $1$ and $2$ for 20 periods of mirror oscillations during the stage II. 
Since Hamiltonian~\eqref{eq:Hions} is a quadratic bosonic theory, exact predictions for the ion quantum simulator can be calculated by solving the linear Heisenberg equations of motion for the annihilation and creation operators, while the results for Moore's model were evaluated numerically by using the method of images discussed in \cite{moore1970quantum}.  

As a function of the driving frequency $\omega_{D}$, one finds peaks of high photon production centered around integer multiples of the frequency $\omega_{1}$.
This finding is in accordance with well known analytical results \cite{dodonov1998resonance}.
The main contribution to these peaks stems from single-mode and two-mode
squeezing, which is connected to the resonance condition $\omega_{D}=\langle\omega_{\ell_{1}}\rangle_{T}+\langle\omega_{\ell_{2}}\rangle_{T}$,
with $\langle\omega_{\ell}\rangle_{T}$ denoting the time average
of the instantaneous eigenfrequency $\omega_{\ell}(t)$ (averaged
over one oscillation period of the mirror). The peaks depicted in
\fref{fig:n_without_correction-1} are slightly shifted from
integer multiples of $\omega_{1}$, because $\omega_{\ell}\neq\langle\omega_{\ell}\rangle_{T}$
and due to artefacts caused by the discretization of the model.

\begin{mdframed}[linecolor=gray]
\begin{center}
  \includegraphics[scale=0.35]{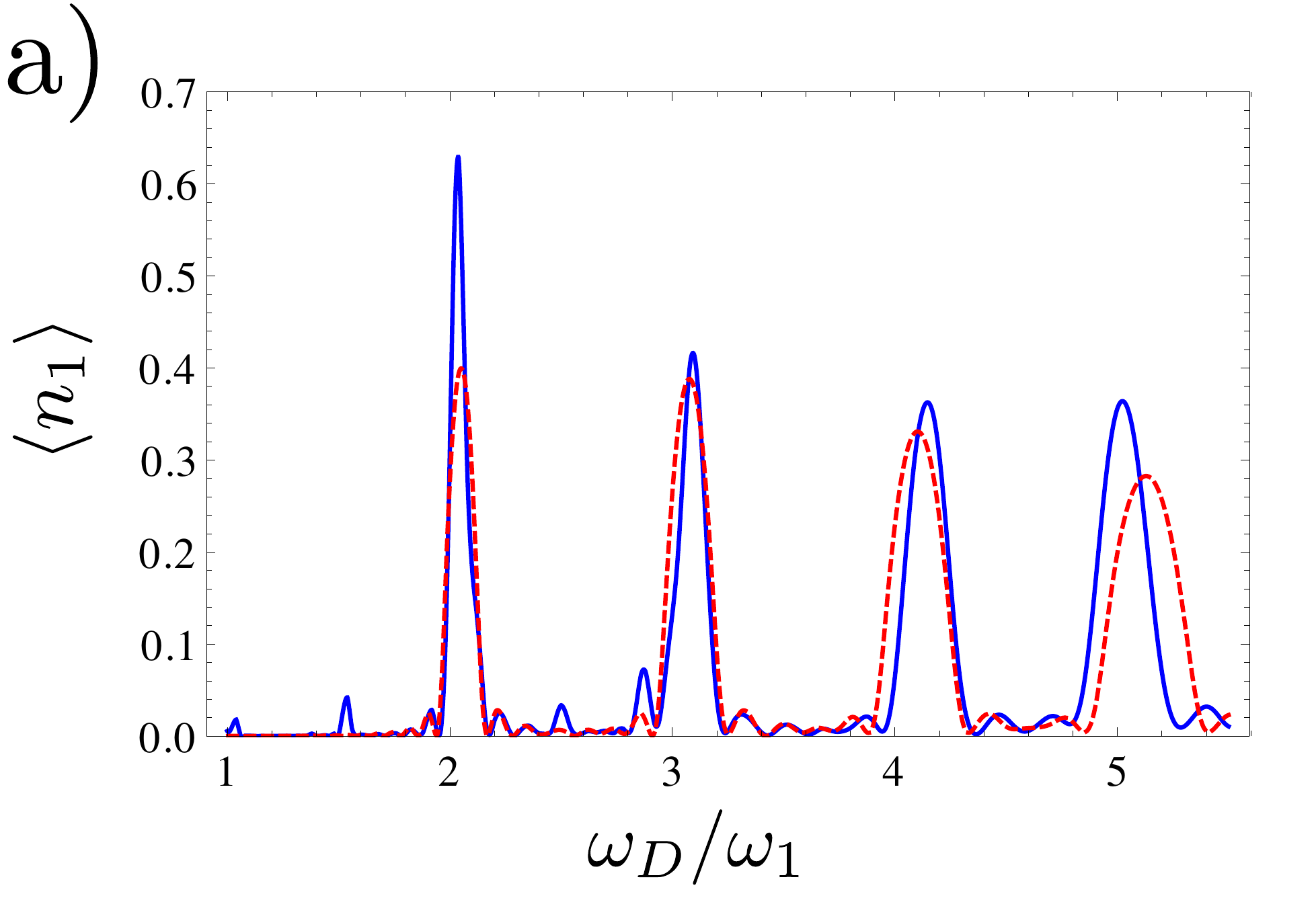} \includegraphics[scale=0.35]{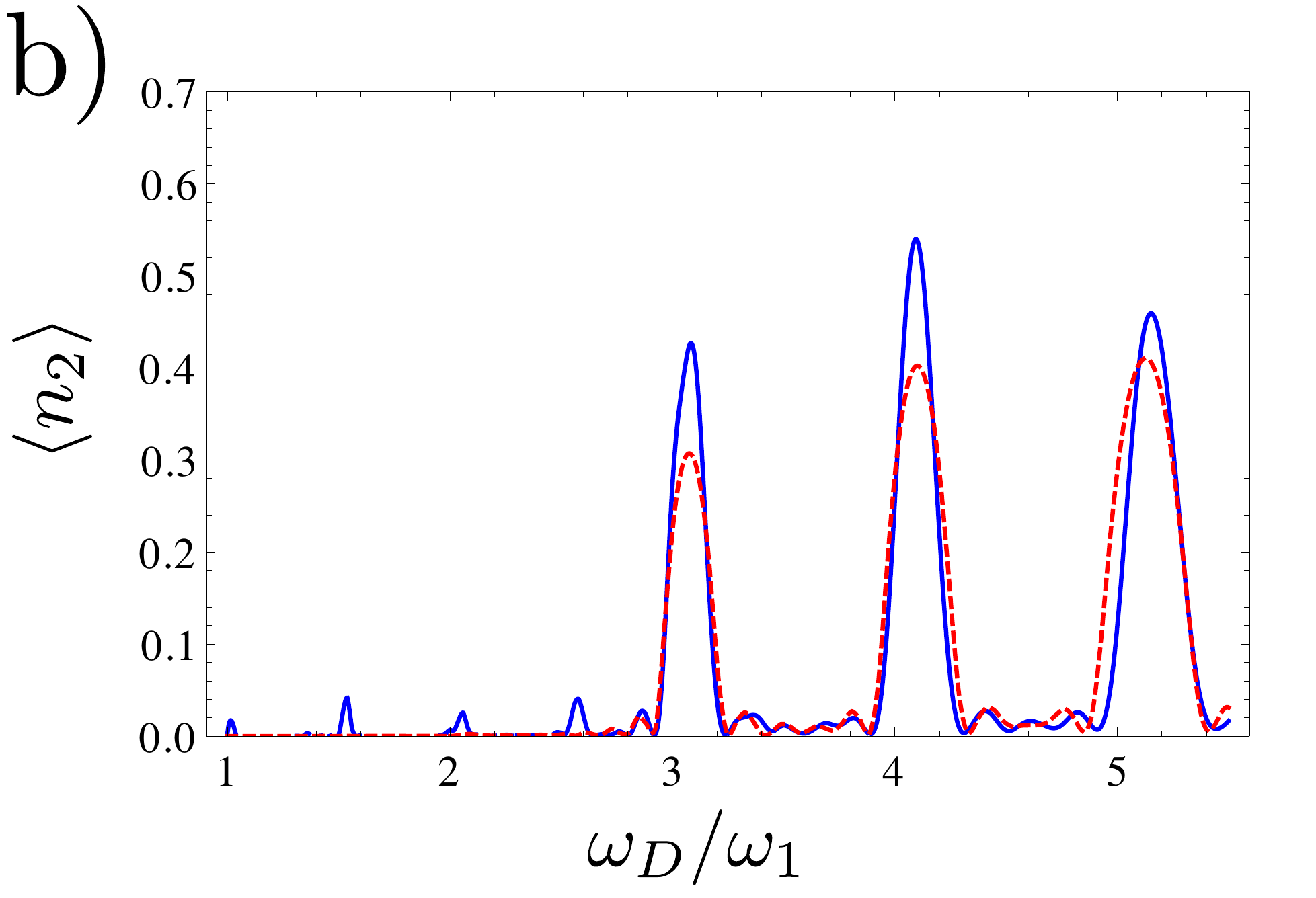}
  \captionof{figure}{\textnormal{Average photon number in (a) the lowest-frequency mode, $\ell=1$, and (b)
the second-lowest frequency mode, $\ell=2$, as a function of the modulation
frequency $\omega_{D}$, evaluated after 20 periods of mirror
oscillation, $\left(t_{2}-t_{1}\right)\omega_{D}=20\cdot2\pi$. Strongly
enhanced photon production is observed at the resonances $\omega_{D}=\langle\omega_{\ell_{1}}\rangle_{T}+\langle\omega_{\ell_{2}}\rangle_{T}$.
Already for the considered small chain of only 20 ions, the expected
results in the ion quantum simulator (blue solid line) show good qualitative
agreement with the ideal results for Moore's model (red dashed line).
\label{fig:n_without_correction-1} }}
\end{center}
\end{mdframed}

Additionally, the trapped-ion quantum simulator allows one to monitor the photon production over time. 
\Fref{fig_Photon_time} displays the corresponding results
for the average photon number in mode 1, for a system driven with
the frequency $\omega_{D}=2\langle\omega_{1}\rangle_{T}$ (with $\langle\omega_{1}\rangle_{T}\approx1.028\omega_{1}$).
For this frequency, which lies is at the first peak in \fref{fig:n_without_correction-1} (a), the photon production is dominated by single-mode squeezing.
When the resonance condition for single-mode squeezing, $\omega_{D}=2\langle\omega_{\ell}\rangle_{T}$, 
is met, the average photon number is approximately given by \cite{dodonov2010current}
\begin{equation}
\langle n_{\ell}\rangle=\sinh^{2}\left[\langle\omega_{\ell}\rangle_{T}\frac{\delta}{4\left(r_{0}-l_{0}\right)}\left(t_{2}-t_{1}\right)\right]\;,\label{eq:approx_resonance_Condition}
\end{equation}
valid if $\delta/\left(r_{0}-l_{0}\right)$ is sufficiently small and the duration of the periodic driving is sufficiently short. At short times,
this approximate expression indeed coincides with the results for
the ion trap as well as for the ideal Moore's model. 
At larger times, the curves start to deviate but the qualitative agreement remains satisfactory.
For reasons of comparison, we also evaluate the phonon production for an improved choice of the time dependence of the optical trapping potential $\omega_{O}^{2}(t)$, which minimizes discretization effects connected to the fact that just a single ion was used to represent the motion of the left mirror. 
Hereby, we chose $\omega_{O}^{2}(t)$ such, that the instantaneous eigenfrequency of the lowest vibrational mode matches the instantaneous eigenfrequency of the lowest optical mode in the idealized cavity setup for all times.
The optimized trajectory lies much closer to the prediction of Moore's model than the simple sine wave, which shows that the deviations are mainly artefacts from using a discretized representation for moving the mirror.

\begin{mdframed}[linecolor=gray]
\begin{center}
  \includegraphics[scale=0.35]{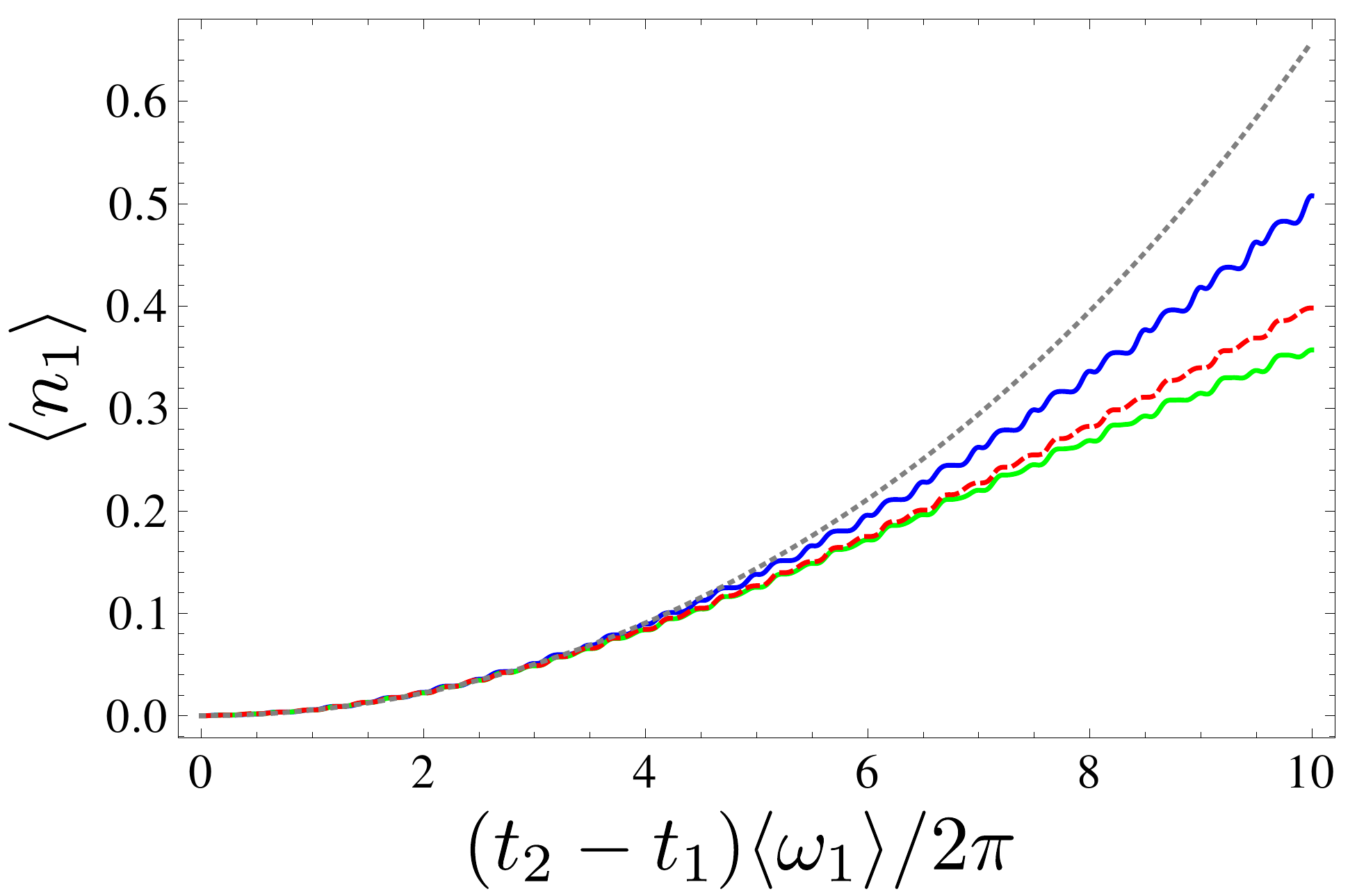}
  \captionof{figure}{\textnormal{Average photon/phonon number in mode 1 plotted over time, for a driving
frequency of $\omega_{D}=2\langle\omega_{1}\rangle_{T}$. The ion-trap
simulation (blue solid line) reaches good qualitative agreement as well with the ideal Moore's model (red dashed line) as with an approximate
analytical result (grey dotted line). 
We also include the prediction for an optimized choice of the temporal dependence of the optical trapping potential $\omega_{O}^{2}(t)$, which minimizes discretization effects (green solid line). 
For short times, the agreement is on a quantitative level. \label{fig_Photon_time} }}
\end{center}
\end{mdframed}

The additional small peaks of the blue curves in \fref{fig:n_without_correction-1} (a) and (b) are also artefacts connected to use of a single ion to represent the motion of the left mirror. 
These artefacts can be reduced by smoothing the mirror motion, i.e., by increasing the number of ions that experience a periodic modulation during stage II. 
Furthermore, by increasing the number of ions representing the radiation field inside the cavity, the distribution of the low-lying eigenfrequencies further approaches the equidistant distribution of the eigenfrequencies in an ideal cavity.
In this way, it is possible to reduce the slight shifts of the peaks seen in \fref{fig:n_without_correction-1} that are caused by discretization effects.

\section{Experimental considerations\label{sec:experimental_considerations}}
In this section, we support our previous  analytical and numerical investigations
by experimental considerations.
We start with a discussion of the robustness of the simulation with respect to
possible sources of errors, such as heating of the ion chain.
Moreover, we review relevant techniques for measuring phononic excitations in ion chains, which allow for the probing of the radiation generated by the DCE.

\subsection{Possible error sources}
As the above results show, an ion chain
with realistic parameters can indeed simulate the photon production
in the DCE, where we find a good agreement to Moore's
model \cite{moore1970quantum} already for 20 ions. Small deviations
do appear due to the limited number of ions. This is no fundamental
limitation, however, and by increasing the number of ions, or by using
additional electrodes, it will be possible to reduce these artefacts
and further improve the simulation of the DCE. As mentioned previously,
the protocol is also rather resilient towards a change in the (time-dependent)
trapping potential. Insufficient control of the spatial dependence
will simply model a slightly modified cavity.

Additionally, in a realistic experiment, one has to make sure that
the observed phonons are not generated by a heating of the ion chain.
In the experimental setup of \cite{vittorini2013modular}, which
has similar parameters to the ones discussed above, a spectral density of electric field
noise of $S_{E}(\omega)\leq3.8\cdot10^{-13}\frac{V^{2}}{m^{2}}\text{Hz}$
has been reported for $\omega=2\pi\cdot1.38\text{MHz}$. Under the
assumption that heating is dominated by electric field noise and that
$S_{E}(\omega)\omega$ is approximately frequency independent, we
obtain a heating rate of the lowest-lying mode [with $\omega_{1}\approx0.21({\overline{k}/m})^{1/2}=2\pi\cdot0.241\text{MHz}$] 
of $1.31\text{ quanta}/\text{ms}$. Hereby, we could neglect the cross
coupling between the RF and noise fields because of the relatively
small frequency $\omega_{1}$. For the higher modes $\omega_{2}$,
$\omega_{3}$, $\dots$, the effect of the electric field noise is
even smaller, since the heating rate it causes (ignoring cross coupling
between RF and noise fields) scales as ${1}/{\omega^{2}}$.
Another source of heating are scattered photons from the laser beam.
The corresponding heating rates can be suppressed by using sufficiently
intense and sufficiently detuned standing-wave laser fields, such
that rates of $0.1\,\text{quanta}/\text{ms}$ seem reasonable \cite{schneider2010optical}.

These heating rates have to be compared to the relevant experimental time scales. 
The data at the first peak in \fref{fig:n_without_correction-1} (a) corresponds to a duration of the experimental stage II of about
$\left(t_{2}-t_{1}\right)=0.041\text{ms}$. Thus, heating is expected
to increase the average phonon number of the 1st mode by roughly $0.054$
phonons, which is one order of magnitude smaller than the number of
phonons generated by the simulated DCE. Even more, one could further
reduce the effect of ion heating due to electric field noise by decreasing
the average nearest-neighbor distance between the ions. This would
increase the frequencies of the modes and hence decrease the time
needed to run the experiment. Therefore, according to these numbers it will
be possible to cleanly observe the first as well as higher peaks in experiment.

\subsection{Probing the radiation field on the singe photon level\label{sec:Probing_radiation}}

As discussed above, the main idea of our ion-chain quantum simulation
of the DCE is to map the photons of the radiation field on the phonons
of the radial ion motion. Thus, for probing the radiation generated
by the DCE, we have to measure the generated phononic excitations in stage III of the experiment.
This can be done with high temporal resolution and high accuracy on the single phonon level by
using the methods available for ion chains \cite{leibfried2003quantum},
which is one of the main advantages of our ion-chain quantum simulation
compared to other schemes \cite{wilson2011observation,dalvit2011quantum,lahteenmaki2013dynamical}.

One possibility for evaluating the number of phonons populating mode
$\ell$ is to drive the corresponding red or blue detuned sideband
for a short time period $\Delta t$ by addressing a single ion with
a laser beam. Hereby, one should choose an ion that takes part in
the collective motion described by the mode $\ell$. By repeating
this experiment for several runs, the probability for exciting the
ion, $P_{e}(\Delta t)$, can be determined. For driving the blue detuned
sideband one obtains 
\begin{equation}
P_{e}(\Delta t)=\sum_{n}p_{\ell}(n)\sin^{2}(\sqrt{n+1}\,\Omega_{0,1}\Delta t)\;,
\end{equation}
where $p_{\ell}(n)$ is the probability of finding $n$ phonons in
mode $\ell$ and $\Omega_{0,1}$ is a characteristic Rabi frequency
determined by the intensity of the applied laser beam and the corresponding
Lamb--Dicke parameter. For short time periods $\Delta t$, the above
expression simplifies to $P_{e}(\Delta t)\approx\Omega_{0,1}^{2}\Delta t^{2}\langle1+\hat{n}_{\ell}\rangle+O(\Delta t^{4})$,
which allows us to evaluate the average phonon number \cite{diedrich1989laser,wineland1987laser}.
By measuring $P_{e}(\Delta t)$ for several $\Delta t$ over a longer
time period, it is possible to determine the phonon number distribution
$p_{\ell}(n)$ by calculating the Fourier transform of $P_{e}(\Delta t)$
\cite{meekhof1996generation,roos1999quantum}. In this way, we can
probe not only the average photon number, but even the detailed photon
statistics of the radiation generated by the DCE.

In principle, it is also possible to apply other methods developed
for probing the quantum state of motion and accessing other observables
\cite{leibfried2003quantum}. The choice of the most suitable method
depends on the experimental parameters of the specific setup.

\section{Conclusions}

In conclusion, we have presented a scheme to realize a quantum simulation
of the DCE in a chain of trapped ions. Hereby, the photons inside
the cavity with moving boundaries are mapped on the phononic excitations
of the radial modes of the ion chain. To achieve the mapping, we derived
a discrete model for the radiation field, which takes the propagation
of radiation within the mirrors into account. We performed a numerical
investigation in which we compared an ion-chain quantum simulation
of the DCE based on realistic experimental parameters with the idealized
model introduced by Moore \cite{moore1970quantum}. Already for $20$
ions, we observe a good quantitative agreement between the ideal realization
of the DCE and our ion trap quantum simulation. The scheme is robust
against the most common sources of errors, and its requirements are
met by many existing experimental ion-trap setups.

The radiation generated by the DCE, including its full statistics,
can be investigated on the single photon respectively phonon level
by using the methods available for ion traps \cite{leibfried2003quantum}.
This possibility of probing the radiation field on the single photon
level is one of the main advantages of our ion-chain quantum simulation
compared to other schemes \cite{wilson2011observation,dalvit2011quantum,lahteenmaki2013dynamical}.
In this article, we mainly focused on the DCE in a 1D cavity with
a single sinusoidally oscillating mirror. It will be interesting to adapt our scheme
to explore further aspects of the DCE, such as the photon production
for non-sinusoidal mirror trajectories \cite{law1994effective,Ugalde2015}, or in  a cavity that oscillates as a whole \cite{lambrecht1996motion},
or the photon production in a semi-infinite system. 
The latter might be realized by simulating a single moving mirror at one side of
the ion chain and by adding a dissipative process \cite{diedrich1989laser} removing the phononic 
excitations from the other side of the chain \cite{givoli1991non,israeli1981approximation}.
The ability to control the confinement of a larger number of ions simultaneously could even
enable us to study the radiation generated by accelerating a single mirror on a non-periodic trajectory,
which can be linked to the Unruh effect and the Hawking radiation emitted by a black hole \cite{davies1977radiation,crispino2008unruh}. 

\section*{Acknowledgements}
Stimulating discussions with J. Marino are acknowledged.
This work is supported by the DAAD, the DFG as part of the CRC 1119 CROSSING, the ERC Synergy Grant UQUAM, EU IP SIQS, and the SFB FoQuS (FWF Project No. F4016-N23).

\appendix

\section{Microscopic model of the Mirrors\label{Appendix_mirrors}}

In this Appendix, we motivate the Hamiltonian $\hat{H}_{1}(t)$,
which models the mirrors. The basic idea is to describe the mirrors according
to the Drude--Lorentz model \cite{rosenfeld1951theory} by a distribution
of charges that can oscillate around fixed positions. These charges
could be the bound electrons of an atom or the electrons in a metal, 
which will oscillate with the plasma frequency. 

The matter--field coupling can be modeled by the interaction term \cite{cohen2001photons}
\begin{equation}
\hat{H}_{\mathrm{i}}=-\int_{\mathbb{R}^{3}}\hat{\mathbf{j}}(\mathbf{x})\cdot\hat{\mathbf{A}}(\mathbf{x})d^{3}\mathbf{x}\;, 
\end{equation}
with being $\hat{\mathbf{j}}(\mathbf{x})$  the charge current. It's Fourier transform
\begin{eqnarray}
 &\hat{\mathbf{j}}_{\omega}(\mathbf{x})=\frac{1}{\sqrt{2\pi}}\int_{\mathbb{R}} \exp\left(i\omega t\right)\hat{\mathbf{j}}(t,\mathbf{x})\text{d}t
\end{eqnarray}
can be connected to the Fourier transform of the electric field
\begin{eqnarray}
 &\hat{\mathbf{E}}_{\omega}(\mathbf{x})=\frac{1}{\sqrt{2\pi}}\int_{\mathbb{R}} \exp\left(i\omega t\right)\hat{\mathbf{E}}(t,\mathbf{x})\text{d}t
\end{eqnarray}
by the relation 
\begin{equation}
\hat{\mathbf{j}}_{\omega}(\mathbf{x})=n(\mathbf{x})\frac{e^{2}}{m}\sum\limits _{m=1}^{N}G_{m}\frac{-i\omega}{\omega_{m}^{2}-\omega^{2}-i\gamma_{m}\omega}\hat{\mathbf{E}}_{\omega}(\mathbf{x})\;,\label{eq:current}
\end{equation}
with the oscillator sum rule $\sum\limits _{m=1}^{N}G_{m}=1$ and $n(\mathbf{x})$ being the
charge density.  
In this relation, the frequencies $\omega_{m}$ and decay rates $\gamma_{m}$ are chosen heuristically to match the experimental findings. 
In the following, we assume that the main contributions in equation \eref{eq:current}
for the relevant frequencies $\omega$ of the radiation field stem
from terms with $\omega_{m},\gamma_{m}\ll\omega$. In this case, we
obtain 
\begin{eqnarray}
 & \hat{\mathbf{j}}_{\omega}(\mathbf{x})=n(\mathbf{x})\frac{ie^{2}}{m\omega}\hat{\mathbf{E}}_{\omega}(\mathbf{x})\\
 & \Rightarrow\hat{\mathbf{j}}(t,\mathbf{x})=-n(\mathbf{x})\frac{e^{2}}{m}\hat{\mathbf{A}}(t,\mathbf{x})\;.
\end{eqnarray}
Hereby, we used that $\hat{\mathbf{E}}(t,\mathbf{x})=-\frac{\partial}{\partial t} \hat{\mathbf{A}}(t,\mathbf{x})$ . The back action of the charge distribution onto the radiation field can be taken into account by the effective Hamiltonian 
\begin{equation}
\hat{H}_{\text{eff}}=\frac{e^{2}}{2m}\int_{\mathbb{R}^{3}}n(\mathbf{x})\hat{\mathbf{A}}^{2}(\mathbf{x})d^{3}\mathbf{x}\;.
\end{equation}
This effective Hamiltonian matches our model of the mirrors, equation~\eref{eq:H_1}.

\section*{References}
  \bibliographystyle{iopart-num}
\bibliography{Literatur_dynamical_Casimir}

\providecommand{\newblock}{}
\begin{thebibliography}{10}
\expandafter\ifx\csname url\endcsname\relax
  \def\url#1{{\tt #1}}\fi
\expandafter\ifx\csname urlprefix\endcsname\relax\def\urlprefix{URL }\fi
\providecommand{\eprint}[2][]{\url{#2}}

\bibitem{Lamb_experiment}
Lamb W~E and Retherford R~C 1947 {\em Phys. Rev.\/} {\bf 72} 241--243

\bibitem{casimir1948attraction}
Casimir H~B~G 1948 On the attraction between two perfectly conducting plates
  {\em Proc. K. Ned. Akad. Wet\/} vol~51 p 150

\bibitem{moore1970quantum}
Moore G~T 1970 {\em J. Math. Phys.\/} {\bf 11} 2679--2691

\bibitem{crispino2008unruh}
Crispino L~C, Higuchi A and Matsas G~E 2008 {\em Rev. Mod. Phys.\/} {\bf 80}
  787

\bibitem{dodonov2010current}
Dodonov V 2010 {\em Phys. Scripta\/} {\bf 82} 038105

\bibitem{johansson2010dynamical}
Johansson J~R, Johansson G, Wilson C~M and Nori F 2010 {\em Phys. Rev. A\/}
  {\bf 82} 052509

\bibitem{johansson2009dynamical}
Johansson J~R, Johansson G, Wilson C~M and Nori F 2009 {\em Phys. Rev. Lett.\/}
  {\bf 103} 147003

\bibitem{wilson2011observation}
Wilson C~M, Johansson G, Pourkabirian A, Simoen M, Johansson J~R, Duty T, Nori
  F and Delsing P 2011 {\em Nature\/} {\bf 479} 376--379

\bibitem{dalvit2011quantum}
Dalvit D~A~R 2011 {\em Nature\/} {\bf 479} 303--304

\bibitem{lahteenmaki2013dynamical}
L{\"a}hteenm{\"a}ki P, Paraoanu G, Hassel J and Hakonen P~J 2013 {\em Proc.
  Natl. Acad. Sci. U.S.A.\/} {\bf 110} 4234--4238

\bibitem{chiaverini2005surface}
Chiaverini J, Blakestad R~B, Britton J, Jost J~D, Langer C, Leibfried D, Ozeri
  R and Wineland D~J 2005 {\em Quantum Inf. Comput.\/} {\bf 5} 419--439

\bibitem{schulz2008sideband}
Schulz S~A, Poschinger U, Ziesel F and Schmidt-Kaler F 2008 {\em New J.
  Phys.\/} {\bf 10} 045007

\bibitem{kaufmann2012thick}
Kaufmann D, Collath T, Baig M~T, Kaufmann P, Asenwar E, Johanning M and
  Wunderlich C 2012 {\em Appl. Phys. B\/} {\bf 107} 935--943

\bibitem{Porras2004b}
Porras D and Cirac J~I 2004 {\em Phys. Rev. Lett.\/} {\bf 93} 263602

\bibitem{Ivanov2009}
Ivanov P~A, Ivanov S~S, Vitanov N~V, Mering A, Fleischhauer M and Singer K 2009
  {\em Phys. Rev. A\/} {\bf 80} 060301(R)

\bibitem{Bermudez2010a}
Bermudez A, Martin-Delgado M~A and Porras D 2010 {\em New J. Phys.\/} {\bf 12}
  123016

\bibitem{Bermudez2011a}
Bermudez A, Schaetz T and Porras D 2011 {\em Phys. Rev. Lett.\/} {\bf 107}
  150501

\bibitem{Pruttivarasin2011}
Pruttivarasin T, Ramm M, Talukdar I, Kreuter A and H{\"a}ffner H 2011 {\em New
  J. Phys.\/} {\bf 13} 075012

\bibitem{Benassi2011}
Benassi A, Vanossi A and Tosatti E 2011 {\em Nat. Commun.\/} {\bf 2} 236

\bibitem{LinDuan2011}
Lin G~D and Duan L~M 2011 {\em New J. Phys.\/} {\bf 13} 075015

\bibitem{Bermudez2013}
Bermudez A, Bruderer M and Plenio M~B 2013 {\em Phys. Rev. Lett.\/} {\bf 111}
  040601

\bibitem{Ramm2014}
Ramm M, Pruttivarasin T and H{\"a}ffner H 2014 {\em New J. Phys.\/} {\bf 16}
  063062

\bibitem{Pyka2013}
Pyka K, Keller J, Partner H~L, Nigmatullin R, Burgermeister T, Meier D,
  Kuhlmann K, Retzker A, Plenio M~B, Zurek W~H, del Campo A and
  Mehlst{\"a}ubler T~E 2013 {\em Nat. Commun.\/} {\bf 4} 2291

\bibitem{Ulm2013}
Ulm S, Rossnagel J, Jacob G, Deg{\"u}nther C, Dawkins S~T, Poschinger U~G,
  Nigmatullin R, Retzker A, Plenio M~B, Schmidt-Kaler F and Singer K 2013 {\em
  Nat. Commun.\/} {\bf 4} 2290

\bibitem{Mielenz2013}
Mielenz M, Brox J, Kahra S, Leschhorn G, Albert M, Schaetz T, Landa H and
  Reznik B 2013 {\em Phys. Rev. Lett.\/} {\bf 110} 133004

\bibitem{Ejtemaee2013}
Ejtemaee S and Haljan P~C 2013 {\em Phys. Rev. A\/} {\bf 87} 051401(R)

\bibitem{delCampo2010}
Del~Campo A, De~Chiara G, Morigi G, Plenio M~B and Retzker A 2010 {\em Phys.
  Rev. Lett.\/} {\bf 105} 075701

\bibitem{leibfried2003quantum}
Leibfried D, Blatt R, Monroe C and Wineland D~J 2003 {\em Rev. Mod. Phys.\/}
  {\bf 75} 281--324

\bibitem{law1994effective}
Law C~K 1994 {\em Phys. Rev. A\/} {\bf 49} 433--437

\bibitem{schneider2010optical}
Schneider C, Enderlein M, Huber T and Sch{\"a}tz T 2010 {\em Nature Photon.\/}
  {\bf 4} 772--775

\bibitem{vittorini2013modular}
Vittorini G, Wright K, Brown K~R, Harter A~W and Doret S~C 2013 {\em Rev. Sci.
  Instrum.\/} {\bf 84} 043112

\bibitem{ospelkaus2011microwave}
Ospelkaus C, Warring U, Colombe Y, Brown K~R, Amini J~M, Leibfried D and
  Wineland D~J 2011 {\em Nature\/} {\bf 476} 181--184

\bibitem{allcock2012heating}
Allcock D~T~C, Harty T~P, Janacek H~A, Linke N~M, Ballance C~J, Steane A~M,
  Lucas D, L J~J~R, Habermehl S~D, Blain M~G {\em et~al.\/} 2012 {\em Appl.
  Phys. B\/} {\bf 107} 913--919

\bibitem{brown2011coupled}
Brown K~R, Ospelkaus C, Colombe Y, Wilson A~C, Leibfried D and Wineland D~J
  2011 {\em Nature\/} {\bf 471} 196--199

\bibitem{stick2006ion}
Stick D, Hensinger W, Olmschenk S, Madsen M~J, Schwab K and Monroe C 2006 {\em
  Nat. Phys.\/} {\bf 2} 36--39

\bibitem{wesenberg2008electrostatics}
Wesenberg J~H 2008 {\em Phys. Rev. A\/} {\bf 78} 063410

\bibitem{meplan1996exponential}
M{\'e}plan O and Gignoux C 1996 {\em Phys. Rev. Lett.\/} {\bf 76} 408--410

\bibitem{dodonov1998resonance}
Dodonov V 1998 {\em J. Phys. A\/} {\bf 31} 9835--9854

\bibitem{diedrich1989laser}
Diedrich F, Bergquist J~C, Itano W~M and Wineland D~J 1989 {\em Phys. Rev.
  Lett.\/} {\bf 62} 403--406

\bibitem{wineland1987laser}
Wineland D~J, Itano W~M, Bergquist J~C and Hulet R~G 1987 {\em Phys. Rev. A\/}
  {\bf 36} 2220--2232

\bibitem{meekhof1996generation}
Meekhof D~M, Monroe C, King B~E, Itano W~M and Wineland D~J 1996 {\em Phys.
  Rev. Lett.\/} {\bf 76} 1796--1799

\bibitem{roos1999quantum}
Roos C, Zeiger T, Rohde H, N{\"a}gerl H, Eschner J, Leibfried D, Schmidt-Kaler
  F and Blatt R 1999 {\em Phys. Rev. Lett.\/} {\bf 83} 4713--4716

\bibitem{Ugalde2015}
Corona-Ugalde P, Martin-Martinez E, Wilson C~M and Mann R~B 2015 {\em arXiv
  preprint arXiv:1511.07502\/}

\bibitem{lambrecht1996motion}
Lambrecht A, Jaekel M~T and Reynaud S 1996 {\em Phys. Rev. Lett.\/} {\bf 77}
  615--618

\bibitem{givoli1991non}
Givoli D 1991 {\em J. Comput. Phys.\/} {\bf 94} 1--29

\bibitem{israeli1981approximation}
Israeli M and Orszag S~A 1981 {\em J. Comput. Phys.\/} {\bf 41} 115--135

\bibitem{davies1977radiation}
Davies P and Fulling S 1977 Radiation from moving mirrors and from black holes
  {\em Proceedings of the Royal Society of London A: Mathematical, Physical and
  Engineering Sciences\/} vol 356 (The Royal Society) pp 237--257

\bibitem{rosenfeld1951theory}
Rosenfeld L 1951 {\em Theory of electrons\/} vol~1 (North-Holland)

\bibitem{cohen2001photons}
Cohen-Tannoudji C, Dupont-Roc J and Grynberg G 2001 {\em Photons and atoms:
  introduction to quantum electrodynamics\/}

\end{thebibliography}
\end{document}